\documentstyle[12pt,epsf]{article}
\newcommand{\req}[1]{(\ref{#1})}
\newcommand{\bel}[1]{\begin{equation}\label{#1}}
\newcommand{\belar}[1]{\begin{eqnarray}\label{#1}}

\begin{document}
\def\epsk{\epsilon_k}
\def\epsj{\epsilon_j}
\def\epsl{\epsilon_l}
\def\om{\omega }
\def\Om{\Omega }
\def\omr{\omega_{rot}}
\def\jx{\hat J_x}
\def\rft{{\tilde \chi}}
\def\mev{\;{\rm MeV}}
\def\respc{\mbox{\boldmath$\chi$}_{\rm coll}(\om)}
\def\bra{\big\langle}
\def\ket{\big\rangle}
\def\mv{\vert m\vert}
\def\mvp{\vert m^{\prime}\vert}
\def\kr{k_{\rho}}
\def\nr{n_{\rho}}
\def\nrp{n^{\prime}_{\rho}}
\def\nnr{N_{\rho}}
\def\nnrp{N^{\prime}_{\rho}}
\def\pr{\prime}
\def\sp{s^{\prime}}
\def\mpr{m^{\prime}}
\title{The Effect of Nuclear Rotation on the Collective
Transport Coefficients
}
\author{F.A.Ivanyuk$^{1,2}$ and  S.Yamaji$^{2}$   \\
\small\it{1) Institute for Nuclear Research, 03028 Kiev, Ukraine, e-mail:
ivanyuk@kinr.kiev.ua}\\
\small\it{2)
RIKEN,
2-1 Hirosawa, Wako-Shi,  Saitama 351-0106, Japan}\\
\small\it{ e-mail: yamajis@rikaxp.riken.go.jp}
}
\maketitle

\begin{abstract}
We have examined the influence of rotation on the potential energy and
the transport coefficients of the collective motion (friction and mass
coefficients). 
For axially symmetric deformation of nucleus $^{224}Th$ we found 
that  at excitations corresponding to
temperatures $T\ge 1~MeV$ the shell correction to the liquid drop energy
practically does not depend on the angular rotation. The friction and mass
coefficients obtained within the linear response theory for the same
nucleus at temperatures larger than $2~MeV$ are rather
stable with respect to rotation {\it provided} that the contributions
from spurious states arising due to the violation of rotation symmetry are
removed. At smaller excitations both friction and mass parameters
corresponding to the elongation mode are growing functions of rotational
frequency $\omr$.
\end{abstract}

\section{Introduction}
\label{sec: intro}
The recent success of Flerov Laboratory, JINR, Dubna in the synthesis of the
superheavy compound systems with $Z=114, 116$ has provoked a considerable
theoretical
interest to the fusion-fission reactions at low excitation energies.
Commonly such reactions are described by solving the Langevin equation
\cite{aberep}-\cite{frorep}
for the collective variables which specify the shape of the nuclear
compound system formed in the result of fusion of heavy ions with
nuclei.
Usually such systems are formed with rather high
angular momentum. The effect of rotation on the fusion or fission
probability is included at most in the calculation of the
macroscopic part of the deformation energy. The possible dependence on
rotation of
the shell correction as well as friction and inertia is
completely ignored.
However one might expect the strong dependence on rotation of the transport
coefficients since the rotation changes considerably the
single-particle spectrum. To the best of out knowledge this effect was
analyzed only in \cite{pomhof}  where rather strong dependence of
friction coefficient on angular velocity was found.

In the present paper we have continued the investigations along the line
presented in \cite{pomhof}. Besides the friction coefficient we have
paid also attention to the rotational dependence of the mass parameter
and the shell correction. The computations are carried out with
two-center shell model \cite{maruhn,suivyaha} which allows for rather
flexible parametrization of the shape around the touching point and
which was used earlier in dynamical computations \cite{arwaoh}. As the
compound nucleus we chose the system
$^{208}Pb+^{16}O\Longrightarrow^{224}Th$ for which the experimental
information on the temperature dependence of the damping parameter is
available \cite{hobapa} and which was studied in \cite{yaivho,ivhopaya} without
account of the rotation.

Due to technical reasons we had to limit ourselves to the
excitations above $T=1 MeV$. The point is that the rotation violates not
only the axial symmetry but the
time reversal symmetry too and the $BCS$ approximation to pairing
interaction breaks down. To account for the pairing accurately one has to
solve a kind of Hartree-Fock-Bogolyubov equation what requires rather
time consuming computation. For this reason we considered here the
excitations corresponding to temperature larger than $T=1 MeV$ when
the pairing can be neglected.

The paper is organized as follows: In Section \ref{sec: forma} we quote the
main relations of the linear response theory adopted for the rotating
nuclei. The quasi--static properties (moment of inertia, liquid drop and
shell component of potential energy) are examined in Section
\ref{sec: static}. The influence of rotation on the response functions and
transport coefficients is investigated in
Secs.\ref{sec: response},\ref{sec: transp}. The
special attention here is paid to the elimination of the contribution
from spurious states caused by rotation. The main conclusions and open
questions are formulated in Summary.
The expressions for the matrix elements of $\hat J_x$ on two-center
oscillator basis wave functions are given in the Appendix.

\section{General formalism}
\label{sec: forma}
By describing the rotating nuclei one usually transforms the Hamiltonian
from the laboratory co-ordinate system to the body fixed (or intrinsic)
co-ordinate system. In the result, instead of the
Hamiltonian $\hat H(Q_{\mu})$ one has to consider the Routhian operator
\bel{routh}
\hat R(Q_{\mu}, \omr)=\hat H(Q_{\mu})-\omr \hat J_x
\end{equation}
with $\omr$ being the rotational frequency and $\hat J_x$ - the
projection of angular momentum on the rotation axes ($x$-axes).
The variables $Q_{\mu}$ in
\req{routh} are the deformation parameters which specify the shape of the
(deformed) mean field.

The energy of rotating nucleus $E=\bra\omr\vert\hat H\vert\omr\ket$ and
angular momentum $I=\bra\omr\vert\hat J_x\vert\omr\ket$ are growing
function of the 
rotational frequency $\omr$. At small values of $\omr$ one can use the
perturbation theory to obtain
\bel{erot}
E=E_0+{1\over 2}{\cal J}_{cran}\omr^2
\end{equation}
where ${\cal J}_{cran}$ is the cranking model moment of inertia. In the
approximation of independent particles it is given by
\bel{mome}
{\cal J}_{cran}=\hbar^2\sum_{kj}{n_k-n_j\over \epsj-\epsk}\vert\bra
k\mid \hat J_x\mid j\ket\vert^2
\end{equation}
The $n_k$ in \req{mome} are the (temperature dependent) occupation
numbers and summation is carried out over the single-particle states $k$
and $j$.
If $\omr$ is not small the single-particle
spectrum $\epsk$ and eigen-functions are to be found numerically by solving
the eigen-values problem,
\bel{eigen}
\hat R(Q_{\mu}, \omr)\vert k \ket =\epsk(Q_{\mu}, \omr)\vert k \ket
\end{equation}

Like in the case without rotation the transport coefficient of collective
motion, the tensors of stiffness $C_{\mu\nu}$, friction $\gamma_{\mu\nu}$
and mass $M_{\mu\nu}$ can be  derived within the linear response theory
\cite{hofrep,yahosa} from  the so called collective response
function $\respc$ approximating $\respc$ by the response function of
damped oscillator
\bel{lorfit}
[{\mbox{\boldmath$\chi$}}_{\rm coll}(\om)]_{\mu\nu}
\longrightarrow [{\mbox{\boldmath$k$}}
(-{\mbox{\boldmath$M$}}\om^2-i{\mbox{\boldmath$\gamma$}}\om+ 
{\mbox{\boldmath$C$}})^{-1}
{\mbox{\boldmath$k$}}]_{\mu\nu}
\end{equation}
Here ${\mbox{\boldmath$k$}}$ is the coupling tensor
, see \req{coupconst} below.
The collective response function  $\respc$ is related to the Fourier
transform $\chi_{\mu\nu}(\om)$ of the intrinsic (causal) response function
\bel{defrest}
\rft_{\mu\nu} (t)= \Theta(t) {i\over \hbar}
             {\rm tr}\,\left(\hat{\rho}_{\rm qs}(Q_{\mu},T)
             [\hat F_{\mu}^{I}(t),\hat F_{\nu}^{I}(0)] \right)
\end{equation}
by
\bel{collresp}
\mbox{{\boldmath $\chi$}}_{\rm coll}(\om)=
\mbox{{\boldmath $\kappa$}}(\mbox{{\boldmath $\kappa$}}+
\mbox{{\boldmath $\chi$}}(\om))^{-1}\mbox{{\boldmath $\chi$}}(\om)
\end{equation}
The ${\mbox{\boldmath$\kappa$}}$ in \req{collresp} is the inverse of
coupling tensor \req{coupconst} and operators $\hat F_{\mu}^{I}(t)$ in
\req{defrest}\ are the interaction representation for the derivatives of
the Routhian $\hat R(Q_{\mu}, \omr)$ with respect to deformation (or
rotational frequency $\omr$),
\bel{generator}
\hat F_{\mu}^{I}(t)=
e^{-{i\over\hbar}\hat R t}\hat F_{\mu}e^{{i\over\hbar}\hat R t},
\quad
\hat F_{\mu}=
{\partial \hat R(Q_{\mu}, \omr)\over\partial Q_{\mu}}
~ , \quad
\hat F_{\omr}=
{\partial \hat R(Q_{\mu}, \omr)\over\partial \omr}
\end{equation}
The average in \req{defrest} is calculated with the quasi-static density
operator for which the canonical distribution is assumed, $\rho_{\rm
qs}(Q_{\mu},\omr,T)\propto exp(-\hat R(Q_{\mu},\omr)/T)$.

The response function \req{defrest} can be used to calculate the
deviation of the average value of $\hat F_{\mu}$ from its quasi-static
value (calculated at some deformation point $Q^0$), (see \cite{hofrep})
\bel{delft}
\delta \bra \hat F_{\mu}\ket_t =
-\sum_{\nu}\int_{-\infty}^{\infty}
\tilde\chi_{\mu\nu}(t-s)(Q_\nu(s) - Q_\nu^0)ds
\end{equation}
The Fourier transform of \req{delft} reads
\bel{delfcoll}
\delta \bra \hat F_\mu\ket_{\om}=
-\sum_{\nu}\chi_{\mu\nu}(\om)\delta Q_{\nu}(\om)
\end{equation}
with $\delta Q_{\nu}(\om)$ being the Fourier transform of $(Q_\nu(s) -
Q_\nu^0)$.

The coupling tensor
${\mbox{\boldmath$k$}}$ (c.f.\cite{hofrep}) is
\bel{coupconst} -\left({\mbox{\boldmath$k$}}^{-1}\right)_{\mu\nu}\equiv
-\kappa_{\mu\nu}= C_{\mu\nu}(0)   +\chi_{\mu\nu}(0)
\end{equation}
where stiffness $C_{\mu\nu}(0)$ of the free energy ${\cal F}(Q,T)$ and static
response $\chi_{\mu\nu}(0)$ are defined by the static properties of the system,
\bel{stiff}
C_{\mu\nu}(0)\equiv
   {\partial^2{\cal F}(Q,T)\over \partial Q_{\mu}\partial Q_{\nu} }
\end{equation}
and $\chi_{\mu\nu}(0)$  is the Fourier transform of the intrinsic
response function \req{defrest} taken at $\om=0$.
\section{Quasi-static properties}
\label{sec: static}
The collective potential energy $E(Q_{\mu},I)$ is one of the most
essential ingredients appearing in the theory of large scale collective
motion.
The derivatives of $E(Q_{\mu},I)$ with respect to deformation define
the collective conservative forces. The second derivatives of
$E(Q_{\mu},I)$ (stiffness) is used to find the inverse of the
coupling tensor $\kappa_{\mu\nu}$ \req{coupconst} which appears in the
collective response function \req{collresp}.

Like in the case without rotation we will use for calculation of the
potential energy the Strutinsky shell correction method
\cite{strut,brdajepastwo}. The idea of applying the Strutinsky
renormalization to the rotational problem was advanced by Pashkevich et
al
\cite{pasfra,nepafr}, see also \cite{anlale}. Following
\cite{nepafr} one can express the intrinsic energy $E(Q_{\mu},I)$ as
\bel{potenergy}
E(Q_{\mu},I)=E_{LDM}(Q_{\mu},I)+\delta R(Q_{\mu},I)
\end{equation}
where $E_{LDM}(Q_{\mu},I)$ is the liquid drop energy of rotating
nucleus and $\delta R(Q_{\mu},I)$ is the shell correction,
\bel{deltar}
\delta R= \sum_k \epsk n\left({\epsk-\lambda\over T}\right)
-\int_{-\infty}^{\infty}n\left({e-\widetilde\lambda\over T}\right)e
\widetilde g(e)de
\end{equation}
In the case of finite temperature instead of the shell correction to the
intrinsic 
energy one has to consider the shell correction to the free energy
$\delta R\Longrightarrow \delta {\cal F} = \delta R - T\delta S$, where
$\delta S$ is the shell correction to the entropy, see \cite{ivahof}.
The energies $\epsk$ in \req{deltar} are to be found by diagonalization of the
shape dependent Routhian \req{routh} and $\widetilde g(e)$ is the
average density of single-particle states, see \cite{strut,brdajepastwo}.

As argued in \cite{nepafr}, $E_{LDM}(Q_{\mu},I)$ can be represented rather
accurately by
\bel{eldm}
E_{LDM}(Q_{\mu},I)=E_{LDM}(Q_{\mu})+I^2/2{\cal J}_{rig}(Q_{\mu})
\end{equation}
Here, ${\cal J}_{rig}(Q_{\mu})$ is the rigid body moment of inertia for the
rotation around $x$-axes and $E_{LDM}(Q_{\mu})$ is the liquid drop energy
of non-rotating nucleus.

In the computations presented below we will use the two-center shell model
\cite{maruhn,suivyaha}
and consider only axially symmetric shapes. Such shapes are
specified mainly by two parameters: the distance $z_0$ between the centers of
left and right oscillator potentials and the parameter $\delta$ which
fix the spheroidal deformation of the "fragments". Below we will consider the
case when deformations of left and right fragments are the same,
$\delta_1=\delta_2=\delta$. Furthermore we will consider here only a
one-dimensional path in the deformation space and define $\delta=\delta(z_0)$
looking for the minimum of the total energy at fixed $z_0$, see \cite{yaivho}.

The comparison of the rigid body and cranking model moment of inertia is
shown in Fig.1. It is seen that both methods give rather
close results (the pairing was neglected). The dependence of
${\cal J}_{cran}$ on the rotation is also not very strong. So, the main source
of rotational dependence of $E_{LDM}(Q_{\mu}, I)$ is the $I^2$ - term in
\req{eldm}.

The left-hand-side part of Fig.2
shows the
rotational dependence of the liquid drop part of deformation energy (we
suppose that spherical shape  corresponds to $Q_{\mu}=0$)
\bel{edef}
E_{LDM}^{def}(Q_{\mu},I)=E_{LDM}(Q_{\mu},I)-E_{LDM}(Q_{\mu}=0,I)
\end{equation}
As it is seen the rotational dependence of the deformation energy is
rather strong. The fission barrier becomes lower due to rotation and
disappears completely at $I\approx 60\hbar$ for the nucleus $^{224}Th$
shown in the figure.
\begin{figure}
\label{Figure1}
\centerline{\epsfxsize=8cm \epsffile{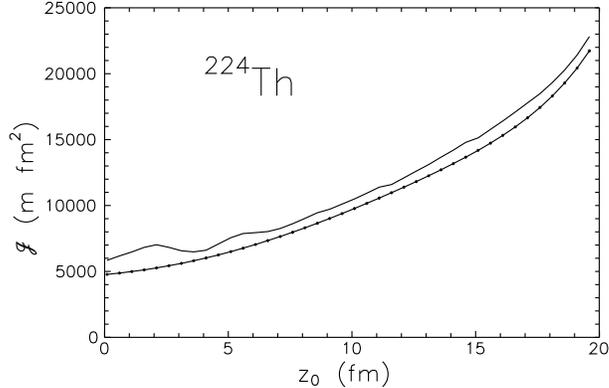}}
\caption
{The deformation dependence of the rigid-body ${\cal J}_{rig.}$
(curve with dots) and the cranking model ${\cal J}_{cran}$ moments of
inertia. The deformation parameter $z_0$ is here the distance between
the centers of left and right oscillator potentials of two-center shell
model and $m$ is the nucleon mass. The ${\cal J}_{cran}$ is computed for
the temperature $T=1~MeV$.}
\end{figure}

The effect of rotation on the fission barriers is known for decades
and taken into account nowadays in all computations of
the deformation energy.
The rotational dependence of the shell correction is less clear. The
diagonalization of Routhian \req{routh} is much more time consuming due
to the break of axial symmetry as compared with the diagonalization of
the non-rotating shell model Hamiltonian. Hence it is assumed usually that this
dependence is weak and the shell correction is computed at
$\omr=0$ only. To clarify this point we have computed the shell
correction for several values of $I$ as a function of deformation
along the liquid drop fission valley of $^{224}Th$. Indeed, see
right-hand-side of Fig.2
, the fluctuation of $\delta {\cal F}$
is less then $1~MeV$ for variation of $I$ from zero to $I=60\hbar$. Very
likely such weak dependence of $\delta {\cal F}$ on $I$ can be neglected.

The weak dependence of the shell correction on rotation is not so
surprising. It was pointed out by Strutinsky \cite{strut87} that the
shell effects are not sensitive to rotation as far as the
perturbation $\hbar\omr$ is small compared with the spacing
$\hbar\Omega_0$ between the gross
shells, $\hbar\omr\ll\hbar\Omega_0$, with $\hbar\Omega_0\approx 8-10 \mev$.
For the maximal value of spin $I=60\hbar$ shown in Fig.2 the
$\hbar\omr=\hbar I/{\cal J}\approx 0.4 \mev$ 
at the saddle and $\hbar\omr\approx 0.6
\mev$ at the minimum. Both values of $\hbar\omr$ are much smaller than 
$\hbar\Omega_0$. Since the moment of inertia $\cal J$ increases with
growing deformation the $\hbar\omr$ gets smaller (for fixed $I$). This
explains, at least partly, why at large deformation the shell correction
is less sensitive to rotation as compared with small deformation, see Fig.2.

\begin{figure}
\label{Figure2}
\centerline{\epsfxsize\textwidth\epsffile{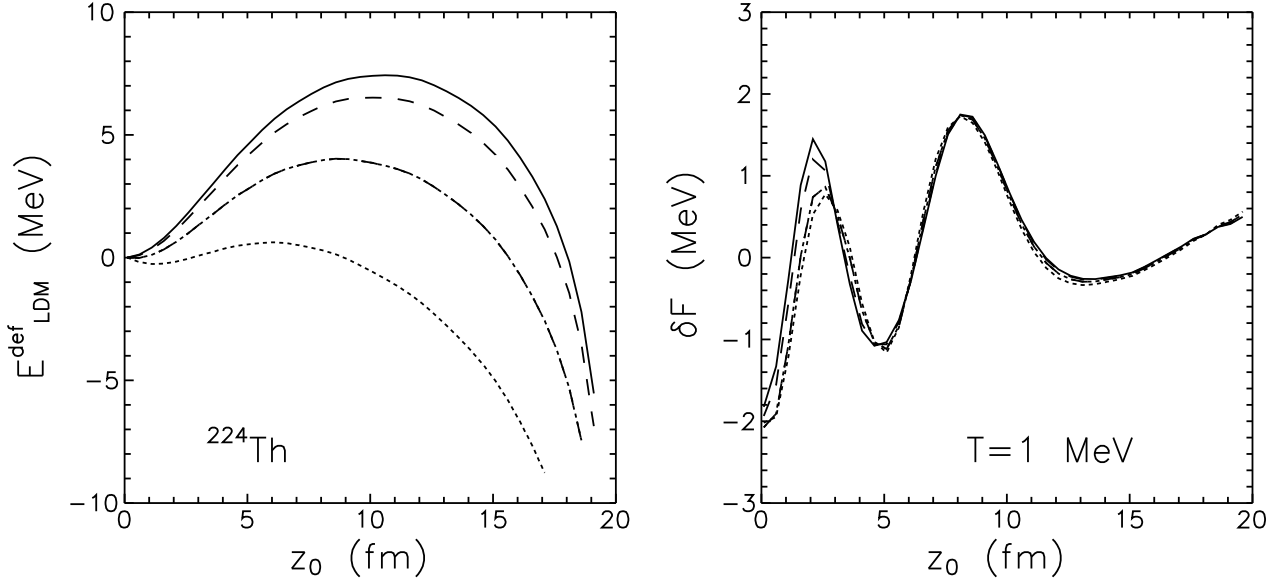}}
\caption
{The liquid drop deformation energy (left) and the shell
correction $\delta {\cal F} =$ $ \delta R - T\delta S$ to the free energy
(right) for temperature $T=1~MeV$ as
function of the deformation parameter $z_0$. The solid, dashed,
dotted-dashed and dotted lines correspond to the values of angular 
momentum equal to 0, 20, 40 and 60$\hbar$.}
\end{figure}
\section{Response functions}
\label{sec: response}
The intrinsic response function $\chi_{\mu\nu}(\om)$ is one of the most
simple and important quantity of the linear response
theory. The friction and mass coefficients in the so called zero
frequency approximation
are expressed in terms of derivatives of the
intrinsic response functions, see \cite{hofrep}. The intrinsic
response function is also an
important ingredient of the collective response function \req{collresp}.
So we will look first at the effect of rotation on the
intrinsic response function.
\subsection{Intrinsic response function}
\label{subsec: intrinsic}
The Fourier transform of the intrinsic response function given by \req{defrest}
can be expressed as the sum over single-particle states
\bel{respfunsum}
\chi_{\mu\nu} (\omega) =\sum_{jk} \chi_{jk}(\omega)
    F^{\mu}_{jk}F^{\nu}_{kj}
\end{equation}
with
\bel{respfun}
\chi_{jk}(\omega) =
-\int_{-\infty}^\infty\;{d\Omega \over 2\pi\hbar }\; n(\Omega )\;
    \Bigl(\varrho_k(\Omega ) {\cal G}_j(\Omega +\omega +i\epsilon)+
    \varrho_j(\Omega ) {\cal G}_k(\Omega - \omega -i\epsilon)\Bigr)
\end{equation}
Here $n(\Omega)$ is the Fermi function determining the occupation of the
(rotation dependent) single-particle levels. The ${\cal G}_k$ appearing in
(\ref{respfun}) is the one-body Green function
\bel{green}
{\cal G}_k(\omega \pm i\epsilon)={1\over{\hbar \omega
    -\epsilon_k-\Sigma^{\prime} (\omega ,T)  \pm i\Gamma(\omega ,T)/2}}
\end{equation}
It is parameterized in terms of the real and imaginary part of the self-energy
$\Sigma(\omega,T)=\Sigma^{\prime}(\omega,T)-i\Gamma(\omega,T)/2$. The
$\Gamma(\omega,T)$ is assumed to have the form
\bel{imselfenomt}
\Gamma(\omega,T)=
    {1\over \Gamma_0}\;{(\hbar \omega - \mu)^2 + \pi^2 T^2 \over
    1 +{\left[(\hbar \omega - \mu)^2 + \pi^2 T^2 \right]/c^2}}
\end{equation}
and $\Sigma^{\prime}(\omega,T)$ is coupled to $\Gamma(\omega,T)$  by
the Kramers-Kronig relation.
The $\varrho_k(\omega)$ represents the
distribution of single-particle strength over more complicated states.
It is related to ${\cal G}_k$ by
\bel{rhogreen}
\varrho_k(\omega)= i({\cal G}_k(\omega +i\epsilon)
-{\cal G}_k(\omega -i\epsilon))
\end{equation}

For the simplified case when the collisional damping could be neglected
the intrinsic response function attains the form
\bel{intresp}
\chi_{\mu\nu} (\omega) =\sum_{kl} {n_k-n_l\over \hbar(\om-\om_{kl})+i0}
    F^{\mu}_{lk}F^{\nu}_{kl}
\end{equation}
with $\hbar\om_{kl}\equiv\epsk-\epsl$ and $i0$ being infinitely small
imaginary constant.

\begin{figure}
\label{Figure3}
\centerline{\epsfxsize\textwidth \epsfysize=0.85\textheight 
\epsffile{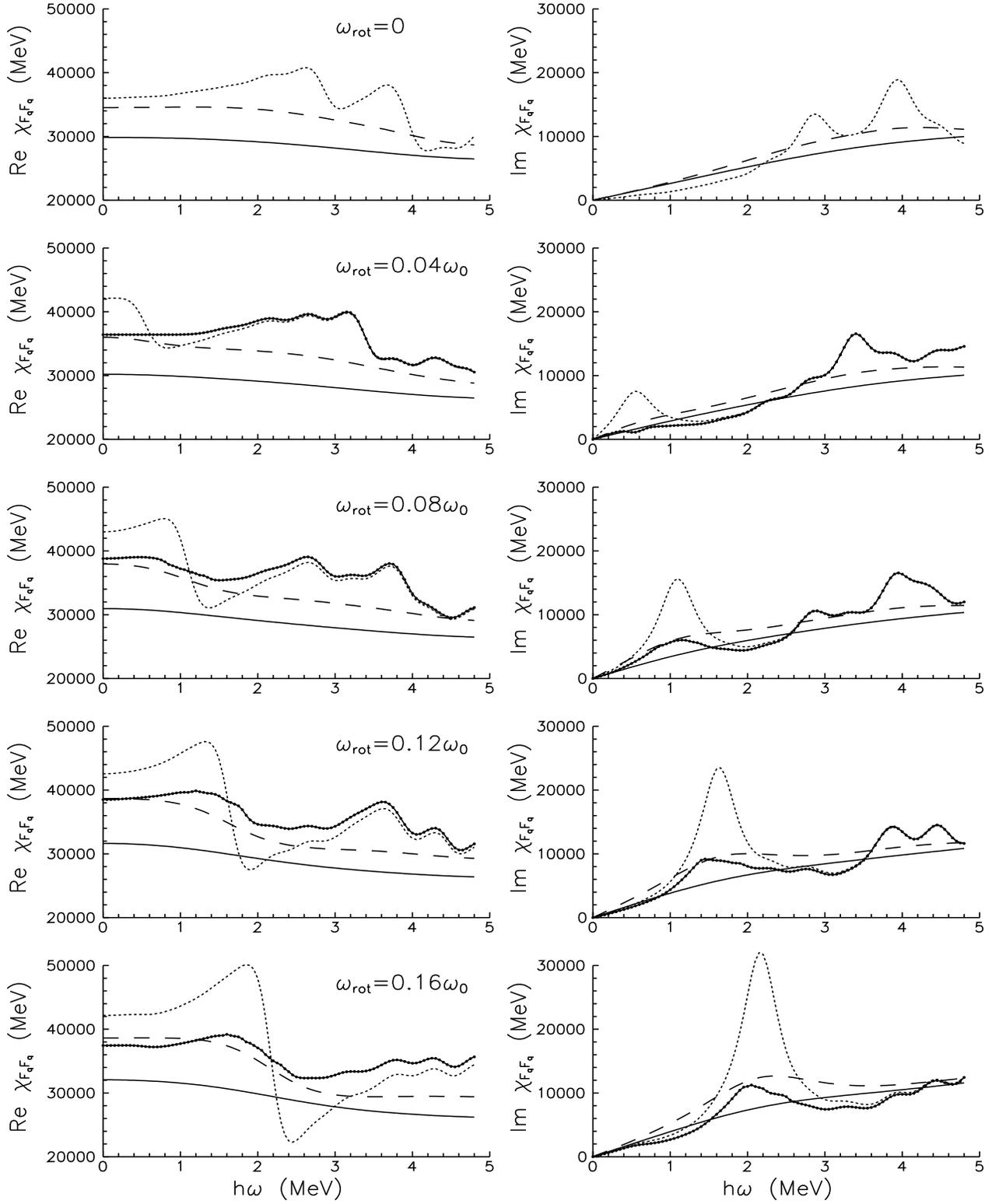}}
\caption
{The frequency dependence of the real (left) and imaginary
(right) parts of $F_qF_q$ intrinsic response
function for several values of the rotational frequency $\omr$. The dotted,
dashed and solid curves correspond to temperatures $T=1,~2$ and $3 \mev$.
The heavy solid curve marks the modified response function
$\widehat\chi_{FF}(\om)$, see \protect\req{modify}, for $T=1 \mev$.
The computations are done for the "ground state" shape of nucleus
$^{224}Th$ very close to the sphere, $z_0=0.1$, parameters of spheroidal
deformations $\delta_1=\delta_2=0.05$. }
\end{figure}

Fig.3 shows few examples of the response function
\req{respfunsum}-\req{rhogreen} for the elongation mode. The deformation
parameter in this case is the distance $q=q(z_0,\delta)$ between the left and
right
centers of mass (divided by the diameter of the sphere with the same
volume) and $\hat F_q$-operator is , see \cite{yaivho}
\bel{foper}
\hat F_{q}\equiv
{\partial \hat H \over \partial z_0}{\partial z_0\over \partial q}+
{\partial \hat H \over \partial \delta}{\partial \delta \over \partial q}
=\left({\frac{\partial q}{\partial z_{0}}+
\frac{\partial \delta}{\partial z_0}
\frac{\partial q}{\partial\delta}}\right)^{-1}
\left(\frac{\partial\hat H}{\partial z_0}+
\frac{\partial \delta}{\partial
z_0}\frac{\partial\hat H}{\partial\delta}\right)
\end{equation}
where the derivative ${\partial \delta}/{\partial z_0}$ is to be
taken along the fission path $\delta=\delta(z_0)$.

Comparing the response function corresponding to different values of $\omr$
one notices a peak in the low frequency region. This peak
is absent in the case $\omr=0$. With growing $\omr$ the peak gets
"stronger" and moves away from $\om =0$. Its position is
approximately proportional to $\omr$. This circumstance hints that this
additional peak may be caused by rotation (let us call it here "rotational"
peak).
Recalling that the friction coefficient $\gamma$ and mass parameter
$M$ (at least in the zero
frequency limit)
are defined by the derivatives of
response function with respect to $\om$ at $\om=0$ it
is clear that the value of both $\gamma$ and $M$ can be very
sensitive to the "rotational" peak. The numerical results show that
the contribution from
"rotational" peak to the real part of the response function can lead to
a {\it negative} value of mass parameter. Thus, the
"rotational" peak could be of spurious origin and one has to treat this
problem very accurately in order to get the reliable results for
friction and inertia.
\subsection{The conservation of angular momentum}
\label{sec: spurio}
We will consider in this section the case of a single deformation parameter
$Q$ for simplicity. The generalization to multi-dimensional case is
straightforward.

It is clear that the Routhian $R(Q, \omr)=\hat H(Q)-\omr \hat J_x$
violates the rotational symmetry.
The  operator $\mbox{\boldmath$J$}$  of angular momentum does not commute
with Routhian, thus $J$ is not a good quantum number. By
varying $\omr$ one can fix the {\it average} value
of $\hat J_x$ , i.e. one chooses $\omr$ in such a way that
\bel{omegarot}
\bra \omr \vert \hat J_x \vert \omr \ket = I
\end{equation}
where by $\vert \omr \ket$ we denote the Irast state of Routhian
\req{routh} calculated for given $\omr$.

If in addition to rotation we switch on also the vibrations
\bel{routhvibra}
\hat H(Q)-\omr \hat J_x\Longrightarrow \hat H(Q^0)-\omr \hat J_x +
\hat F \delta Q(t)
\end{equation}
then $\vert
\omr \ket$ becomes time-dependent, $\vert \omr \ket \Longrightarrow
\vert \omr \ket_t$ and, in principle, average value of $\jx$ is not
conserved any more
\bel{deltai}
\bra \omr \vert \jx \vert \omr \ket_t = I(t)=
I+\delta \bra\jx \ket_t  \neq I
\end{equation}
The variation $\delta \bra\jx \ket_t$ is given within the linear
response theory by \req{delft}, namely
\bel{deljxt}
\delta \bra\jx \ket_t=
-\int_{-\infty}^{\infty}
\tilde\chi_{J_x F}(t-s)(Q(s) - Q^0)ds
\end{equation}
The $\delta \bra\jx \ket_t$ given by \req{deljxt} is not zero.
The possible way to make $\bra \omr \vert \jx \vert \omr \ket_t$
time independent is to allow $\omr$ to dependent on time
\belar{deltao}
\omr &\Longrightarrow& \omr(t)=\omr^0+\delta \omr(t)\nonumber\\
\hat H(Q)-\omr \jx&\Longrightarrow& \hat H(Q^0)-\omr^0 \jx +
\hat F \delta Q(t) -\jx \delta \omr (t)
\end{eqnarray}
The time-dependent correction $\delta \omr(t)$ should be found from the
requirement that $\delta \bra\jx \ket_t$ (or its Fourier transform
$\delta \bra\jx \ket_{\om}$) is equal to zero,
\bel{require}
\delta \bra\jx \ket_t = \delta \bra\jx \ket_{\om}=0
\end{equation}
This problem can be easily solved by means of linear response theory.
Considering the time-dependent part $\hat F\delta Q(t)
-\jx\delta\omr(t)$ as a small perturbation one can find (mind \req{delft})
\belar{deljxt1}
\delta \bra\jx \ket_t=
-\int_{-\infty}^{\infty}\tilde\chi_{J_x F}(t-s)(Q(s) - Q^0)ds\nonumber\\
-\int_{-\infty}^{\infty}\tilde\chi_{J_xJ_x}(t-s)(\omr(s) - \omr^0)ds,
\end{eqnarray}
or its Fourier transform
\bel{deltaj}
\delta \bra\jx \ket_{\om}=-\chi_{J_x F}(\om)\delta Q(\om)-
\chi_{J_xJ_x}(\om)\delta \omr(\om)
\end{equation}
The analogous expression can be also written for the variation $\delta
\bra F \ket_{\om}$
\bel{deltaf}
\delta \bra F \ket_{\om}=-\chi_{FF}(\om)\delta Q(\om)-
\chi_{FJ_x}(\om)\delta \omr(\om)
\end{equation}
Recalling \req{require} one can find $\delta \omr(\om)$ from \req{deltaj} as
\bel{deltaom}
\delta \omr(\om) = - \chi_{J_x F}(\om)\delta Q(\om)
/\chi_{J_xJ_x}(\om)
\end{equation}
In principle, $\delta \omr(t)$ could be found by Fourier transform
of \req{deltaom}. But this is not necessary for practical purpose. One
can insert $\delta \omr(\omega)$ in \req{deltaf} to define the modified
response function $\widehat \chi_{FF}(\om)$,
\bel{deltaf2}
\delta \bra F \ket_{\om}=-\widehat\chi_{FF}(\om)\delta Q(\om)
\end{equation}
with
\bel{modify}
\widehat \chi_{FF}(\om)=
\chi_{FF}(\om)-{\chi_{FJ_x}(\om)\chi_{J_x F}(\om)\over\chi_{J_xJ_x}(\om)}
\end{equation}
The modified response function \req{modify} is shown by heavy solid line
in Fig.3. It is seen that "rotational peak" has disappeared.
Consequently, the  transport coefficients computed
with modified response function \req{modify} will not contain the
contributions from the "rotational peak" and would differ considerably from
these derived with $\chi_{FF}(\om)$.

The above method was successfully used in \cite{ivahof} to remove the
contributions to the response function caused by the violation of the
particle number conservation by pairing. It was demonstrated there that
fixing of the average value of the particle number with the
time-dependent density matrix leads to the same secular equation for the
vibrational mode as obtained within RPA. This method is rather general
and can be used to fix any physical quantity. For example, using
transformation analogous to \req{modify} one can remove the center of
mass motion in case of isoscalar dipole vibrations.

It is of certain interest to compare the secular equation which results
from the above approach with one obtained earlier, for example within the so
called "cranked RPA" \cite{mikign,janmik}.

It is argued in
\cite{mikign,janmik} that for the description of non-rotational
excitations in the rotating nuclei one should substitute the Routhian
\req{routh} by some supplementary rotational invariant Hamiltonian
$\widetilde H$
\bel{wideh}
\hat R=\hat H-\omega_{rot} \hat J_x \Longrightarrow
\widetilde H=\hat H-h(\hat{\mbox{\boldmath$J$}}^2)
\end{equation}
The many-body Hamiltonian $\hat H$ was approximated in
\cite{mikign,janmik} by the mean-field part $\hat H_0$ plus
quadrupole-quadrupole interaction
\bel{hspheric}
\hat H=\hat H_0-{\kappa\over 2}\sum_{m=-2}^{m=2}(-1)^m\hat Q_{2m}
\hat Q_{2-m}
\end{equation}
\bigskip
and for $h(\hat{\mbox{\boldmath$J$}}^2)$ the expansion was used
\bel{hrotinv}
h(\hat{\mbox{\boldmath$J$}}^2)=\bra h \ket+\omega_{rot} 
(\hat J_x-\bra \hat J_x \ket)
+\mu_x(\hat J_x-\bra \hat J_x \ket)^2
+\mu(\hat J_y^2+\hat J_z^2)+ ...
\end{equation}
with
\bel{mumux}
\omega_{rot}=2\mu \bra \hat J_x \ket,\qquad \mu_x={1\over 2}{d^2\bra\hat H\ket/
d\bra\hat J_x\ket^2}
\end{equation}
The terms omitted in \req{hrotinv} are of the higher order in $\hat J_x-\bra
\hat J_x \ket$, $\hat J_y$ or $\hat J_z$.
Solving of the Hamiltonian $\widetilde H$ within RPA leads to the secular
equation
\bel{mihosec}
\omega^2{\cal F}^+(\omega)=0
\end{equation}
with
\bel{fplus}
\begin{array}{ccccccc}
&\Biggr\arrowvert&S_{xx}&S_{x0}&S_{x2}&\Biggr\arrowvert&\\ {\cal F}^+(\omega) =
&\Biggr\arrowvert&S_{0x}&S_{00}-\kappa/2&S_{02}&\Biggr\arrowvert&\\
&\Biggr\arrowvert&S_{2x}&S_{20}&S_{22}-\kappa/2&\Biggr\arrowvert&
\end{array}
\end{equation}
where
\bel{sij}
S_{\mu\nu}=\sum_{k>l}\left\{
{(n_k-n_l)\omega_{kl}\over\omega^2-\omega_{kl}^2}
q^{(\mu)}_{kl}q^{(\nu)}_{kl}+
{(n_{\bar k}-n_{\bar l})\omega_{\bar k\bar l}\over
\omega^2-\omega_{\bar k\bar l}^2}
q^{(\mu)}_{kl}q^{(\nu)}_{kl}\right\}
\end{equation}
The spurious zero energy state is separated by \req{mihosec} and all
other excitations are given by ${\cal F}^+(\omega)=0$.

To get the equation for the collective frequencies in our approach
let us recall that the equation
\req{require} together with the self-consistency condition
\bel{selfcons}
\delta \bra \hat F \ket_{\om}=k\delta Q(\om)
\end{equation}
leads to the system of equations (mind \req{deltaf},\req{deltaj})
\belar{system}
(k+\chi_{FF}(\om))\delta Q(\om)+\chi_{FJ_x}(\om)\delta \omr(\om)=0
\nonumber\\
\chi_{J_x F}(\om)\delta Q(\om)+\chi_{J_xJ_x}(\om)\delta \omr(\om)=0
\end{eqnarray}
The eigenfrequencies for the system \req{system} are found from the
equation
\bel{determin}
{\rm Det}(\om)\equiv (k+\chi_{FF}(\om))\chi_{J_xJ_x}(\om)-
\chi_{FJ_x}(\om)\chi_{J_x F}(\om) = 0
\end{equation}
If one would neglect the effects of collisional damping then one could
use expressions \req{intresp} for the response functions. In this case
${\rm Det}(\om)$ coincides exactly with the nontrivial part $\cal F^+(\om)$
of the secular equation \req{mihosec} obtained in \cite{mikign} (one
should also put $S_{2\nu}=S_{\mu 2}=0$ since only axially symmetric shapes
are considered in present work). So Eq.\req{determin} does not contain
the spurious contributions caused by the violation of rotational symmetry.
\begin{figure}[ht]
\label{Figure4}
\centerline{\epsfxsize\textwidth\epsffile{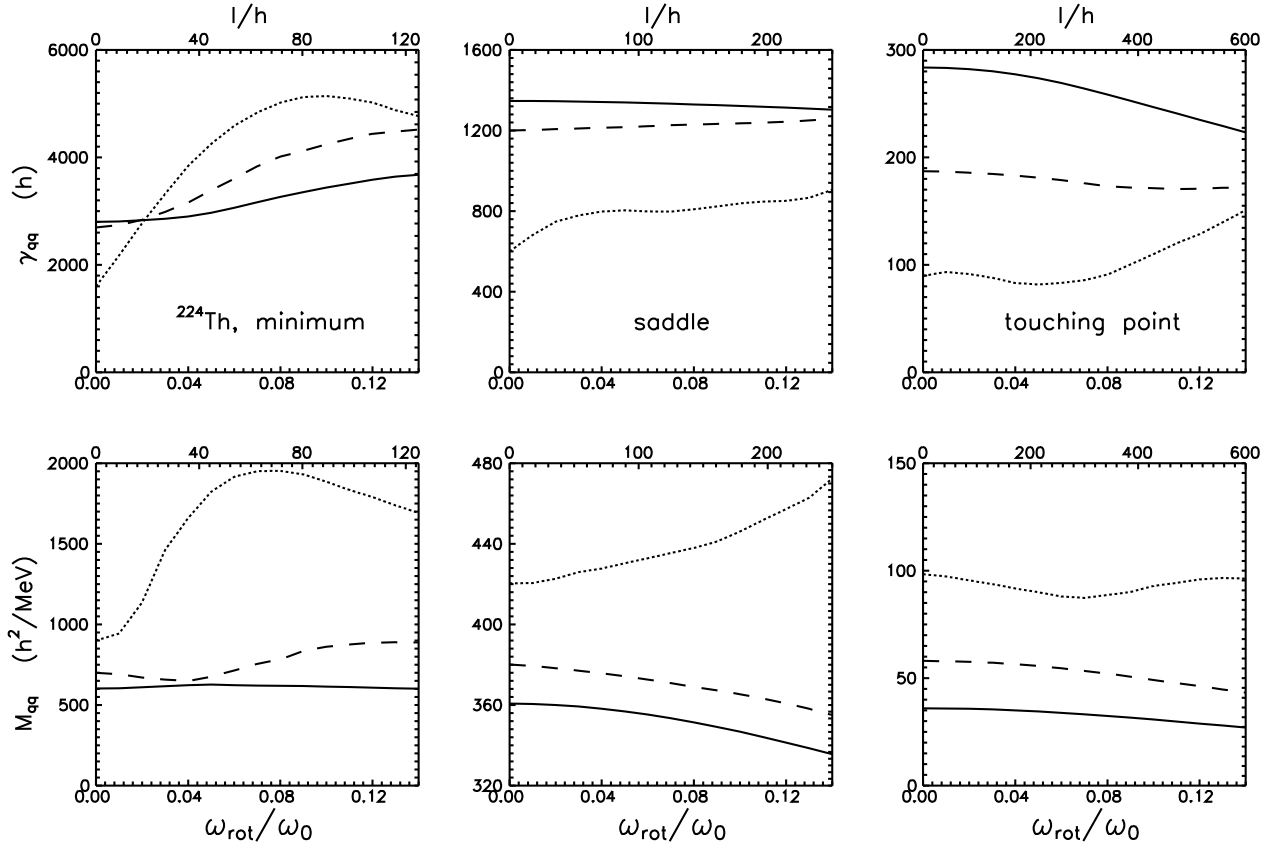}}
\caption
{The friction and inertia \protect\req{lorfit} for the
elongation mode as functions of the average value of angular momentum $I$
(upper $x$-axes) or rotational frequency $\omr$ (lower $x$-axes).
Dotted, dashed and solid curves correspond to temperatures $T=1, 2$ and
$3~MeV$. The computations are done for several deformations of nucleus
$^{224}Th$ which correspond to the minimum of potential energy, saddle
and two touching spheres.}
\end{figure}
\section{Transport coefficients}
\label{sec: transp}
In Fig.4 we show the friction $\gamma_{qq}$ and mass $M_{qq}$ 
coefficients defined by the fit
\req{lorfit} of the collective response function
for three particular deformations which are of a special
interest: at the ground state deformation of $^{224}Th$, at the saddle
and for at touching point. The last configuration is of the interest
for description of initial stage of fusion reactions. The
deformation parameter $q\equiv R_{12}/2R_0$ is here the distance
$R_{12}$ (divided by the diameter of nucleus) between the centers of
mass of left and right parts of nucleus. The advantage of such a choice is
explained in \cite{yaivho,ivhopaya}. As it is seen from
Fig.4 the dependence of $\gamma_{qq}$ on $\omr$ is much
weaker as that found in \cite{pomhof}. Evidently, this is because we
have removed the spurious "rotational" peak from the response function.
Without modification \req{modify} we would obtain the friction
coefficient which look very much like that of \cite{pomhof}.

One can see from Fig.4 that for the ground state deformation and
temperatures $T=1 MeV$ the friction and mass coefficients depend
somewhat on the rotation. Both $\gamma_{qq}$ and
$M_{qq}$ increase with $\omr$ in the interval $0 \leq \omr\leq
0.08\om_0$. For higher $T$ and $\omr$ both $\gamma_{qq}$ and
$M_{qq}$ are rather stable with respect to variation of $T$ and
$\omr$. For more deformed shapes the friction and mass coefficients are
not very sensitive to the rotation for all temperatures.
\begin{figure}[htp]
\label{Figure5}
\centerline{\epsfxsize\textwidth\epsffile{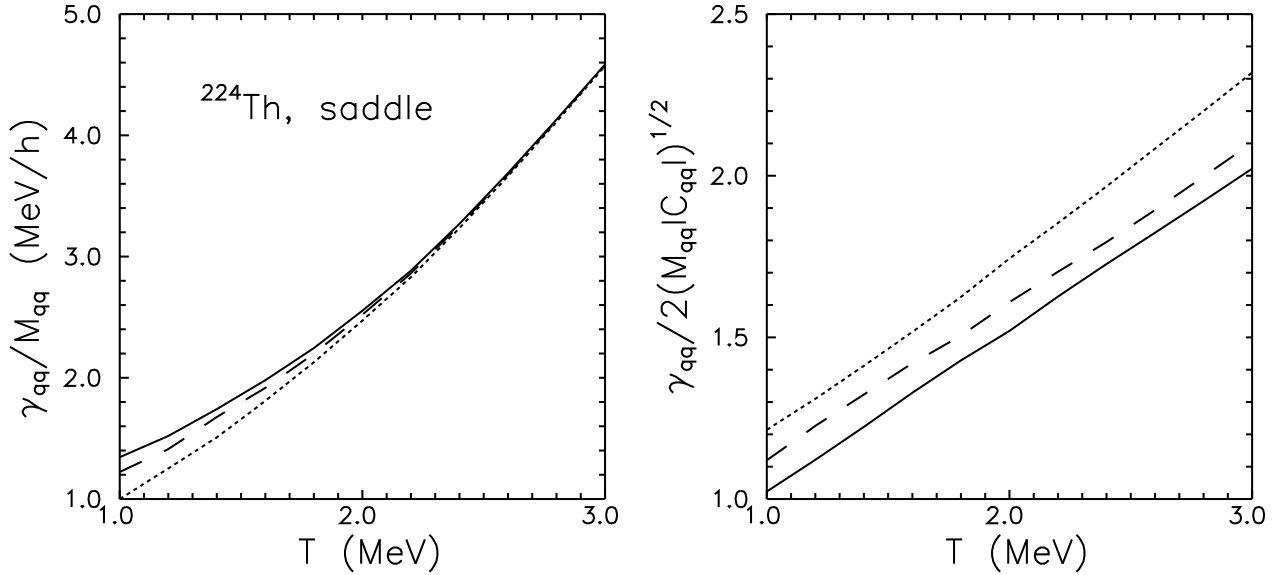}}
\caption
{The reduced friction coefficient
$\beta_{qq}=\gamma_{qq}/M_{qq}$ (left)
and the damping
factor $\eta_{qq}=\gamma_{qq}/2\protect\sqrt{\vert C_{qq}\vert M_{qq}}$
(right) versus temperature. The dotted,
dashed and solid curves correspond to the values of angular momentum
equal to 0, 40 and 60 $\hbar$.} 
\end{figure}

Finally, Fig.5 shows the reduced friction coefficient
$\beta_{qq}=\gamma_{qq}/M_{qq}$ and the damping
factor $\eta_{qq}=\gamma_{qq}/2\protect\sqrt{\vert C_{qq}\vert M_{qq}}$
at the saddle of $^{224}Th$ as the function of temperature. The damping factor
reveals whether collective motion is underdamped ($\eta<1$) or overdamped
($\eta>1$). As Fig.5 shows, the collective motion changes
from underdamped to overdamped at $T\approx 1 MeV$.
Both $\beta_{qq}$ and $\eta_{qq}$ shown in Fig.5 increase with the
temperature. This behaviour is in a qualitative agreement with the one
found in \cite{hobapa}. The increase of $\eta_{qq}$ with the temperature is
impossible to explain neither with the wall friction nor with the
hydrodynamical viscosity. The increase of
$\eta_{qq}$ with temperature obtained here is not as rapid as that found
in \cite{hobapa}. The account of rotation does not diminish this
discrepancy since both $\beta_{qq}$ and $\eta_{qq}$ shown in Fig.5
do not depend much on the rotation. The dependence of $\eta_{qq}$ on $I$
seen from Fig.5 is mainly due to some dependence of liquid drop
stiffness on rotation. We should also note that in this work we did not
normalize the high temperature limit of the 
mass parameter to the irrotational flow value. That is
why the numerical results for the mass parameter and, consequently,
reduced friction $\beta_{qq}$ and damping factor $\eta_{qq}$ differ
somewhat from these obtained in \cite{yaivho}.
\section{Summary and outlook}
\label{subsec: summa}
We have examined the influence of rotation on the transport
coefficients of the collective motion.  
Rather unexpectedly we have found out that friction $\gamma$
and mass $M$ parameters for rotating nuclei are rather sensitive to such
fine effects as the violation of rotational symmetry by Coriolis term
$-\omr \hat J_x$. For the ground state deformation 
the spurious contributions to collective friction and
mass are (at least) as large as  those of physical importance.
This circumstance was
not clear (to the best of our knowledge) up to now.

In order to remove the spurious contributions we had to  modify the
model of "stationary rotation" and to introduce the time-dependent
rotational frequency. In the result we obtained the friction and the mass which
demonstrate rather reasonable dependence on the rotational frequency
$\omr$. For excitations above $T=2 MeV$ when the microscopic shell
effects disappear both $\gamma$ and $M$ are rather insensitive to the
rotation, i.e. behave like macroscopic quantities.  For $1 MeV\leq T
\leq 2 MeV$ we found some increase of $\gamma$ and $M$ with growing
$\omr$. Such effect might be caused by the change of shell structure due
to the re-arrangement of single-particle states by rotation.

Even stronger dependence of $\gamma$ and $M$ on $\omr$ should be
expected for $T\leq 1 MeV$ when both shell end pairing effects are
present.  As it was shown in \cite{ivahof,hofiva} the pairing effects
change considerably the collective transport at low excitations. The
destruction of pairing by the rotation can have
considerable effect on the transport coefficients and is worth to be
examined. Such details could be very important for the accurate
description, for example, of the final stage of the fusion reaction and
formation of superheavy elements which takes place at low excitation
energy.

The extension of the method developed in the present work to the
simultaneous treatment of both pairing and rotation will be the subject
of future work.

\bigskip
{\bf Acknowledgements}. The authors are grateful to H.Hofmann who
has drawn their attention to the problem of the dependence of transport
coefficients on the angular rotation as well as to I.N.Mikhailov,
V.V.Pashkevich and K.Pomorski for fruitful
discussions. One of us (F.A.I) would like to thank the Cyclotron Center,
RIKEN for the hospitality extended to him during his stay in Japan.

\appendix
\section{The matrix elements of ${\jx}$ operator}
\label{sec: waves}
The single-particle basis wave functions are defined in the two-center
shell model \cite{maruhn,suivyaha} as
\bel{spbasis}
\vert n_z n_{\rho}m~ s_z\ket=\varphi_{n_z}(z)~ \chi^{\vert m
\vert}_{n_{\rho}}(\rho)~ \eta_m(\varphi)~\chi_{1/2}(s_z)
\end{equation}
where
\bel{etam}
\eta_m(\varphi)={1\over\sqrt{2\pi}}~e^{im\varphi}~,
\end{equation}
with $m$ being an arbitrary integer number and
\bel{chirho}
\chi^{\mv}_{n_{\rho}}(\rho)=(-1)^{{m+\mid m \mid \over 2}}
\sqrt{{2(n_{\rho})~!}\over{(n_{\rho}+{\mv})~!}}
~k_{\rho}^{{\mv+1}\over 2}
~e^{-{1\over 2}\kr \rho^2}~\rho^{\mv}~L^{\mv}_{\nr}(\kr \rho^2)
\end{equation}
Here $\kr={m_0\omega_{\rho} /\hbar}$, $\nr$ is a not-negative integer and
$L^{\mv}_{\nr}(\xi)$ is a Laguerre polynomial.
The $z-$components of wave function
\bel{phizet}
\varphi_{n_z}(z) =
\cases{~N^{-1}_{n_{z1}}~U(-n_{z1}-{1\over 2}, -\sqrt{2k_{z1}}(z-z_1))
&for~$z<0$~~~\cr
~N^{-1}_{n_{z2}}~U(-n_{z2}-{1\over 2},~~~ \sqrt{2k_{z2}}(z-z_2))
&for~$z>0$~~~\cr}
\end{equation}
 are expressed in terms of
parabolic cylinder functions $U(a,x)$,
\belar{uax}
U(a,x) =
{\sqrt{\pi}\over2^{{a/ 2}+{1/4}}}~
{_1F_1({a/2}+{1/4},{1/2},{x^2/2})
\over \Gamma({a/2}+{3/4})}e^{-x^2/4}-\nonumber\\
{\sqrt{2\pi}x\over2^{{a/ 2}+{1/4}}}{_1F_1({a/2}+{3/4},{3/2},{x^2/2})
\over \Gamma({a/2}+{1/4})} e^{-x^2/4}
\end{eqnarray}
The constants $N_{n_{z1}}$ and $N_{n_{z2}}$ are defined by the
normalization and the continuity of $\varphi_{n_z}(z)$ and its first
derivative at $z=0$, see \cite{maruhn}.

The operator of single-particle angular momentum $\hat{\bf j}$ is
\bel{jot}
\hat{\bf j}=\hat{\bf l}+\hat{\bf s}
\end{equation}
where $\hat{\bf l}$ is the orbital momentum 
$\hat{\bf l}=-i~[{\bf r}~{\bf \nabla}]$
and $\hat{\bf s}$ is the spin
$\hat{\bf s}={1\over 2}~\mbox{\boldmath$\sigma$}$,
with $\mbox{\boldmath$\sigma$}$ being the Pauli matrices. 
From \req{jot} it follows immediately that
\bel{matrix}
\bra n_z n_{\rho}m~ s_z \vert \hat j_x\vert n^{\pr}_z
n^{\pr}_{\rho}\mpr~ \sp_z\ket
=\bra n_z n_{\rho}m~\vert \hat l_x\vert n^{\pr}_z n^{\pr}_{\rho}\mpr~\ket
\delta_{s_z,\sp_z}+
\bra s_z \vert \hat s_x\vert \sp_z\ket\delta_{n_z,n_z^{\pr}}
\delta_{n_{\rho},n_{\rho}^{\pr}}\delta_{m,\mpr}
\end{equation}
For the spin part of \req{matrix} it is easy to find
\bel{spinmat}
\bra s_z\vert \hat s_x \vert s^{\pr}_z\ket=
{1\over 2}(\delta_{\sp_z,s_z+1}+\delta_{\sp_z,s_z-1})
\end{equation}
The $x$-component of the orbital
momentum $\hat l_x$ is given in the cylindrical co-ordinate system
$\{r,\theta,\varphi\}$ by
\bel{lxcyl}
\hat l_x=i\left\{\left(z{\partial\over{\partial\rho}}-
\rho{\partial\over{\partial z}}\right)\sin\varphi+
{z\cos\varphi\over\rho}{\partial\over{\partial\varphi}} \right\}
\end{equation}
Note that $\hat l_x^*=-\hat l_x$.

If the nucleus is not left-right symmetric then the rotation axes does
not go through $z=0$ but through the center of mass $z_{cm}$.
Consequently eq.\req{lxcyl} should be modified to
\bel{lxcylcm}
\hat l_x=i\left\{\left((z-z_{cm}){\partial\over{\partial\rho}}-
\rho{\partial\over{\partial z}}\right)\sin\varphi+
{(z-z_{cm})\over\rho}\cos\varphi{\partial\over{\partial\varphi}} \right\}
\end{equation}
Since the single particle wave functions are separable in
$\{\rho,z,\varphi\}$
the matrix elements \\
$\bra n_z n_{\rho}m~\vert l_x\vert n^{\pr}_z n^{\pr}_{\rho}\mpr~\ket$
are then the product of one-dimensional matrix elements
\belar{matrix1}
\bra n_z n_{\rho}m~\vert l_x\vert n^{\pr}_z n^{\pr}_{\rho}\mpr~\ket=
\nonumber\\
\bra n_{\rho}m\vert\rho\vert  n^{\pr}_{\rho}\mpr\ket
\bra m\vert -i\sin\varphi\vert \mpr\ket
\bra n_z \vert {\partial\over \partial z}\vert n^{\pr}_z\ket +
\bra n_z \vert z-z_{cm}\vert n^{\pr}_z\ket\times\nonumber\\
\left[\bra n_{\rho}m\vert{\partial\over{\partial\rho}}\vert
n^{\pr}_{\rho}\mpr\ket\bra m\vert i\sin\varphi\vert \mpr\ket+
\bra n_{\rho}m\vert {1\over\rho}\vert n^{\pr}_{\rho}\mpr\ket
\bra m\vert i\cos\varphi{\partial\over\partial\varphi}\vert\mpr\ket\right]
\end{eqnarray}
The matrix elements $\bra n_z \vert z-z_{cm}\vert n^{\pr}_z\ket$ and
$\bra n_z \vert {\partial/\partial z}\vert n^{\pr}_z\ket$ are the same as
computed in the
two center shell model code \cite{maruhn,suivyaha}.
For $\bra m\vert -i\sin\varphi\vert \mpr\ket$
and $\bra m\vert i\cos\varphi{\partial/\partial\varphi}\vert \mpr\ket$
it is easy to find
\belar{phimat1}
\bra m\vert -i\sin\varphi\vert \mpr\ket=
{1\over 2}(\delta_{\mpr,m+1}-\delta_{\mpr,m-1}),\nonumber\\
\bra m\vert i\cos\varphi{\partial\over\partial\varphi}\vert \mpr\ket=
{-\mpr\over 2}(\delta_{\mpr,m+1}+\delta_{\mpr,m-1})
\end{eqnarray}
What is left to calculate are $"\rho"$-matrix elements. These can be
calculated using the recurrence relation between Laguerre polynomials
and their derivatives. After somewhat lengthy derivation one can find
\belar{final}
\bra n_z n_{\rho}m~\vert \hat l_x\vert n^{\pr}_z n^{\pr}_{\rho}\mpr~\ket=
\nonumber\\
{1\over 2}\left[\sqrt{\nr+m}~\delta_{\mpr,m-1}+\sqrt{\nr}
~\delta_{\mpr,m+1}\right]
\left[\sqrt{k_{\rho}}~\bra n_z\vert z-z_{cm} \vert n^{\pr}_z\ket+
{1\over \sqrt{k_{\rho}}}~
\bra n_z\vert  {\partial\over\partial z}\vert n^{\pr}_z\ket\right]
\delta_{\nnrp,\nnr-1}+\nonumber\\
{1\over 2}\left[\sqrt{\nrp+\mpr}~\delta_{\mpr,m+1}+\sqrt{\nrp}
~\delta_{\mpr,m-1}\right]
\left[\sqrt{k_{\rho}}~\bra n_z\vert z-z_{cm} \vert n^{\pr}_z\ket]-
{1\over \sqrt{k_{\rho}}}~
\bra n_z\vert  {\partial\over\partial z}\vert n^{\pr}_z\ket\right]
\delta_{\nnrp,\nnr+1}
\end{eqnarray}

Here we have introduced the quantum number $N_{\rho}=2n_{\rho}+|m|$. The
expression \req{final} is valid for $m$ and $\mpr$ being {it both}
non-negative.  The matrix elements for non-positive $m$ and $\mpr$ can
be related to
\req{final} using \req{etam} and symmetry properties of $\hat l_x$
\bel{ss5}
\bra n_z n_{\rho}-m~\vert l_x\vert n^{\pr}_z n^{\pr}_{\rho}-\mpr~\ket=
\bra n_z n_{\rho}m~\vert l_x\vert n^{\pr}_z n^{\pr}_{\rho}\mpr~\ket,
\end{equation}
The operator $\hat j_x$ couples the states with $\Delta j_z=\pm 1$. In
this way the states with positive and negative  $j_z$ are coupled to
each other. One can reduce the dimension of matrix to be diagonalized by
factor two using so called Goodman transformation \cite{good}. It was
suggested in \cite{good} to introduce the basis states of the type
\bel{goodst}
\mid K \ket={1\over\sqrt{2}}(\mid k \ket+\mid \bar k \ket), \qquad
\mid \bar K \ket={1\over\sqrt{2}}(\mid \bar k \ket-\mid k \ket)
\end{equation}
where $\mid k \ket=\mid n_zn_{\rho}ms_z\ket$ for such $m$ and $s_z$ that
$m+s_z-1/2\equiv j_z-1/2$ is even. The single particle states
$\mid n_zn_{\rho}ms_z\ket$ with $j_z-1/2$ - odd up to a sign factor
coincide with $\mid \bar k \ket$. It was shown in \cite{good} that the
matrix elements $\bra K \vert \hat j_x \vert \bar K^{\pr}\ket$ are zero 
and the matrix of $\hat j_x$ on the states $\mid K \ket$ is of
quasi-diagonal form. The nonzero matrix elements are
\bel{nonzero}
\bra K \vert \hat j_x\vert K^{\pr}\ket=-
\bra \bar K \vert \hat j_x\vert \bar K^{\pr}\ket=
\bra k \vert \hat j_x\vert \bar k^{\pr}\ket
\end{equation}
So the quantities of interest are matrix elements
$\bra k\vert\hat j_x \vert\bar k^{\pr} \ket\equiv
\bra k\vert\hat j_x T\vert k^{\pr} \ket$.
It turns out possible to express the matrix elements
$\bra K \vert \hat j_x\vert K^{\pr}\ket$ in terms of
$\bra k\vert\hat j_x \vert k^{\pr} \ket$,
\bel{goodfin}
\bra K \vert \hat j_x \vert K^{\pr}\ket=
~\bra n_z n_{\rho}ms_z\vert \hat j_x \vert n^{\pr}_z
n^{\pr}_{\rho}\mpr\sp_z\ket
\end{equation}
For $\bra n_z n_{\rho}ms_z\vert \hat j_x \vert n^{\pr}_z
n^{\pr}_{\rho}\mpr\sp_z\ket$ the expressions
\req{matrix},\req{spinmat},\req{final} are to be used.

The particular case of $m+s_z=\mpr+\sp_z=1/2$ should be considered
separately. In this case one has to combine \req{final} with \req{ss5}
to obtain
\belar{ss9}
\bra n_z,n_{\rho},1/2-s_z,s_z\vert j_x T\vert
n^{\pr}_z,n^{\pr}_{\rho},1/2-\sp_z,\sp_z \ket={1\over 2}
\delta_{n_z,n^{\pr}_z}\delta_{n_{\rho},n^{\pr}_{\rho}}\delta_{\sp_z,1-s_z}
\nonumber\\
+{1\over 2}\delta_{\sp_z,-s_z}\delta_{\nnrp,\nnr-1}
\sqrt{\nr+1/2-s_z}
\left[\sqrt{k_{\rho}}~\bra n_z\vert z-z_{cm}\vert n^{\pr}_z\ket+
{1\over \sqrt{k_{\rho}}}~
\bra n_z\vert  {\partial\over\partial z}\vert n^{\pr}_z\ket\right]
\nonumber\\
+{1\over 2}\delta_{\sp_z,-s_z}\delta_{\nnrp,\nnr+1}
\sqrt{\nrp+1/2-\sp_z}
\left[\sqrt{k_{\rho}}~\bra n_z\vert z-z_{cm}
\vert n^{\pr}_z\ket-
{1\over \sqrt{k_{\rho}}}~
\bra n_z\vert  {\partial\over\partial z}\vert n^{\pr}_z\ket\right]
\end{eqnarray}

\end{document}